\documentstyle[11pt,newpasp,twoside]{article}
\def\edcomment#1{\iffalse\marginpar{\raggedright\sl#1\/}\else\relax\fi}
\reversemarginpar

\begin{document}
\title{Optical Spectroscopy of T dwarfs}
\author{Adam J.\ Burgasser}
\affil{California Institute of Technology, MSC 103-33, Pasadena, CA 91125}
\author{J.\ Davy Kirkpatrick}
\affil{Infrared Processing and Analysis Center, MSC 100-22, Pasadena, CA 91125}

\begin{abstract}
Optical spectra are presented for a sample of T dwarfs, obtained using the
Keck 10m Low Resolution Imaging Spectrograph.
Dominant features are discussed, including possible temperature
diagnostics.  
\end{abstract}

We have used the Low Resolution Imaging Spectrograph (LRIS; Oke et al.\
1995), mounted on the Keck 10m Telescope, to obtain red optical
spectra of nine
T dwarfs selected from Burgasser et al.\ (1999, 2000a, 2000c,
2001), Strauss et al.\ (1999), and Tsvetanov et al.\ (2000).  Observations
of 2MASS 1237+65, SDSS 1346-00, and SDSS 1624+00 are presented in
Burgasser et al.\ (2000b), which discusses instrumental 
setup and data reduction;
the remaining targets, discussed here, were observed on 5 March 2000
(UT).  

Reduced spectra are shown in Figure 1a, normalized at 9200 {\AA} and 
vertically offset by 0.5.  
The rapid rise in flux from
8000 to 10000 {\AA} is likely due to the extremely pressure-broadened K I
doublet at 7665 and 7699 {\AA}, as noted in SDSS 1624+00 by 
Liebert et al.\ (2000).  
A slight rise in flux between 6500 and 7800 {\AA}
is seen in a few objects, most clearly in
2MASS 0559-14 and SDSS 1624+00 (Figure 1b), and is
due to an opacity window between
the K I doublet and the similarly broadened Na I D lines (5890 and 5896 {\AA};
Reid et al.\ 2000).
We note that the increase in flux on the red side of the K I doublet
is greater in Gl 570D (T$_{eff}$ $\sim$ 750 K) than in
2MASS 0559-14 (T$_{eff}$ $\sim$ 1200 K).  
H$_2$O absorption at 9200 {\AA}
similarly increases from 2MASS 0559-14 to Gl 570D.  
The behavior of the Cs I
lines (8521 and 8943 {\AA})
is unclear due to poor signal-to-noise in these faint dwarfs, as well as the
influence of CH$_4$ at 8900 {\AA}; similarly,
Rb I (7948 {\AA}) is detected only in the high signal-to-noise spectrum of
2MASS 0559-14.  These alkali features are seen to strengthen throughout the
L dwarf sequence (Kirkpatrick et al.\ 1999).
FeH (9896 {\AA}), which weakens in mid- to late-L dwarfs,
is seen in 2MASS 0559-14,
SDSS 1624+00, and 2MASS 1225-27, but is not readily apparent in 
the other objects. 
Finally, the H$\alpha$ emission reported in 2MASS 1237+65
(Burgasser et al.\ 2000b) is not seen in any of the other T dwarfs,
in agreement with activity relations observed in late-M 
and L dwarfs (Gizis et al.\
2000), and further evidence that 2MASS 1237+65 may have an unusual activity
mechanism.
We propose that FeH, H$_2$O, and the red slope of the K I doublet provide
potential temperature diagnostics in the optical regime.

\begin{figure}
\plotfiddle{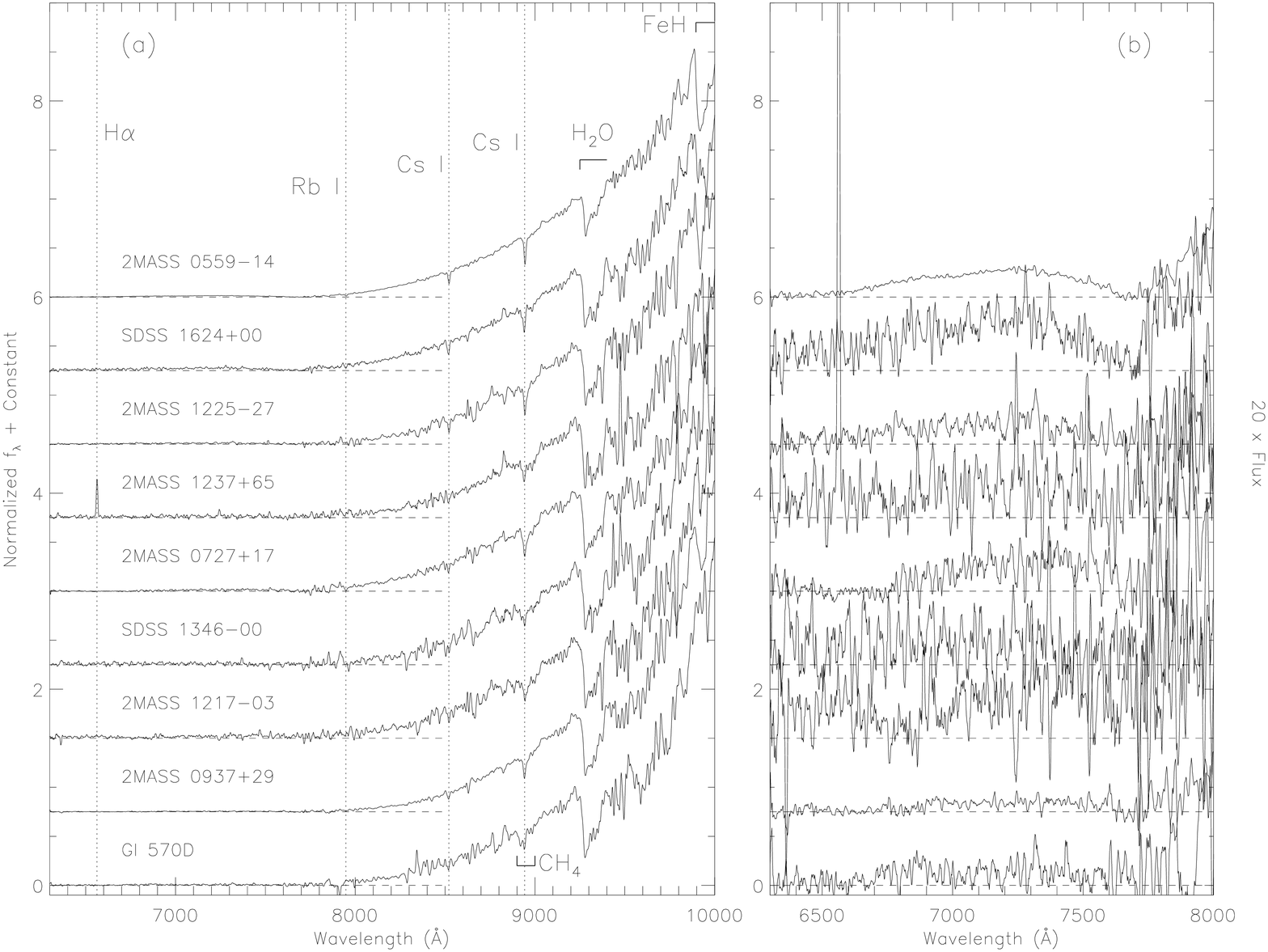}{3.3in}{0}{30}{30}{-175}{-15}
\caption{Optical spectra of T dwarfs. (a) Full 6300--10100 {\AA} spectra,
with dominant features indicated.  
(b) 6300--8000 {\AA} region, magnified 20$\times$
to show flux between Na I and K I doublets.}
\end{figure}

\end{document}